# Microstructural and Superconducting Properties of $YBa_2Cu_{3-x}Co_xO_{7-\delta}$ System

Sushant Gupta[1], R. S. Yadav[1], N. P. Lalla[2], G. D. Verma[3] and B. Das[1]

[1] Department of Physics, University of Lucknow, Lucknow–226007, India
[2] UGC-DAE Consortium for Scientific Research, Indore Centre, Indore-452017, India
[3] Department of Physics, Indian Institute of Technology, Roorkee- 247667, India
Email: bdas226010@gmail.com

## Abstract

Bulk superconductor samples of $YBa_2Cu_{3-x}Co_xO_{7-\delta}$ with x = 0, 0.01, 0.03, 0.05 are synthesized by solid-state reaction route. Both x-ray diffraction and electron microscopy have been employed to study the phase identification, intergrowths, dislocations and the local structure of these samples. Transition temperature of the samples has been determined by four probe resistivity measurements. The x-ray diffraction patterns indicate that the gross structure/ phase of $YBa_2Cu_{3-x}Co_xO_{7-\delta}$ do not change with the substitution of Co up to x=0.05. The zero resistance critical transition temperature [$T_c(R=0)$] is found to decrease and critical current density ($J_c$) increases with the increased concentration of cobalt in the compound. The $J_c$ enhancement for the cobalt doped samples may be resulting due to flux pinning from some defects such as planar defects, stacking faults and micro defects (twin, domains etc.) and the rapid suppression in $T_c$ may be due to the cooper pair breaking and the hole filling in the $CuO_2$ planes.

# 1. Introduction

Chemical doping and introduction of defects in bulk high-Tc superconductors have generated great interest because they represent easily controlled and efficient tool for improving the superconducting properties. $YBa_2Cu_3O_{7-\delta}$ (Y:123 for brevity) is an attractive material for investigation due to their varying critical transition temperature with oxygen content and strong flux pinning capability in high magnetic field. It is well known that the oxygen content affect the crystal structure, electron/ hole transport and superconducting properties in Y:123. It is also realized that the superconducting transition temperature (Tc) sensitively depends on both the hole concentration in the $CuO_2$ planes and the relative concentration of the oxygen within the planes [1]. The level of this concentration can be controlled either by manipulating the oxygen stoichiometry in the Cu-O chains, by application of pressure or by ionic substitution [2, 3].

Several studies of cation substitution of each element constituting the Y:123 compound have been reported in the literature. Their effects on structural and superconducting properties depend on the cation nature and the substitution site. Among all the cationic sites, Cu sites are thought to be of prime importance as superconductivity is primarily supposed to reside in the $CuO_2$ planes. For substitutions at Cu sites, 3d transition metal elements possess certain favorable features due to their compatible ionic sizes and closer orbital structures to that of copper [4, 5]. Further, incorporation of magnetic 3d transition elements provides an interesting prospect of the possible interplay of magnetism and superconductivity occurring in the doped materials. Numerous studies of 3d transition metal ion substitution such as Co, Ni or Zn for Cu in the high temperature superconducting cuprates have shown that Tc is more sensitive to the paramagnetic impurity content than to non-magnetic impurities [6-8]. The substitution of magnetic ion such as Fe at Cu-site in $YBa_2Cu_3O_{7-\delta}$ was studied by Awana et al. [9]. They found that the estimated magnetic moment of Fe provides the possibility of cooper pair breaking as a source of $T_c$ degradation. The process of the moment formation of the Fe seems to correlate indirectly with the mechanism of $T_c$ degradation.

On the other hand many studies have indicated that the chemical doping can improves the flux pinning behavior and mechanical properties of Y:123. As a consequence of both the unavoidability and desirability of defects in the oxide superconductors, studies of these defects are important. With this goal in mind, the effect of the partial substitution of Co at Cu-site, on superconducting properties and micro structural features of $YBa_2Cu_3O_{7-\delta}$ by transmission electron microscopy (TEM) technique have been investigated. Transmission electron microscopy in both imaging and diffraction modes are extremely powerful techniques to determine the local structural characteristics; i.e., to detect deviations from the average structure, as determined by powder x-ray diffraction.

The central aim of the present work is to analyse the physical properties (Transition temperature $T_c$ and Critical current density $J_c$), structure and micro structural changes due to the substitution of Co at Cu site.

## 2. Experimental Details

Samples with nominal composition $YBa_2Cu_{3-x}Co_xO_{7-\delta}$ (where x = 0, 0.01, 0.03, 0.05) were synthesised by standard solid state reaction method. The appropriate ratio of the constituent oxides or carbonate i.e. $Y_2O_3$ (99.9%, Alfa Aesar), CuO (99.99%, Alfa Aesar), $Co_3O_4$ (99.9%, Alfa Aesar) and $BaCO_3$ (99.9%, Alfa Aesar) were thoroughly mixed and ground for several hours (2 to 4 hrs) with the help of mortar and pestle. After regrinding and mixing, the powder was kept in a platinum or alumina crucible and heated (calcined) at 925 $^0$C. The calcinations step served to decompose the carbonate and starting material to interdiffuse for phase formation and chemical homogeneity. After calcinations the material was again ground to subdivide any aggregated products and to further enhance chemical homogeneity. These steps were repeated 3 to 4 times for better homogeneity and phase purity.

The homogeneous powder thus formed was converted into form of pellets before sintering. For this we employed the most widely used technique i.e. dry pressing, which consists of filling a die with powder and pressing at 400 Kg/cm$^2$ into a compacted disc shape. In this way several pellets with varying thickness (1mm to 3mm) were prepared. Finally these pellets were put into alumina crucibles and sintered at about 925$\pm$ 10 $^0$C in air. The heating rate to the sintering temperature was about 100$^0$C/hour. After sintering at 925$\pm$10 $^0$C final annealing was carried out in oxygen atmosphere at partial pressure ~10$^{-1}$ atm, at temperature 550 $^0$C for 13 hr in order to maximize the incorporation of oxygen. The sintering and annealing conditions employed for all the samples are given in Table 1. The annealing temperature is compromised between higher temperatures where oxygen diffusion rates are higher and lower temperatures where the thermodynamic equilibrium favours higher oxygen contents. Some important factors that can determine 'Sample quality' during sample preparation are the stoichiometry, the mixing procedure, the sintering temperature, the oxygen flow during annealing and the rate of sample cooling. If the sample preparation is done with care, a better understanding of the roles these major factors play should ensure that final samples are of high quality.

The structure and phase purity of the powder sample ground from sintered pellets were examined by x-ray diffraction technique using a powder x-ray diffractometer (18 KW, Rigaku Japan) with CuK$_\alpha$ radiation, $\lambda$ = 1.5408Å. The diffraction data were collected over the diffraction angle range of 2θ= 0-90$^0$ by step scanning with a scanning rate 2$^0$/minute. The surface morphology of as synthesized materials has been carried out by a Joel scanning electron microscope (JSM-5600) operated at 25 kV, with a resolution power of 3.5nm. The structural/microstructural characteristics was explored by transmission electron microscope (TEM, Tecnai 20 $^2$G) in both the imaging and diffraction modes.

The electrical resistivity has been measured by standard four probe method in the temperature range 20-300 K using Keithley source meter (2400); Keithley nanovoltmeter (2182A) and Lake shore temperature controller (330). Samples are cooled in liquid helium cryogenic system and heated 0.5 $^0$K/minute. The samples used for resistivity measurements have dimensions of about 0.2×0.14×0.3cm$^3$ and the connection of copper leads with the sample is made using silver paint. Finally in order to see the effect of Co doping on the physical properties, the transport critical current density (J$_c$) value has been measured by standard four-probe method as potential difference of 1μv/cm appears across the sample by increasing current at temperatures ranging from 77 to 60 K.

## 3. Results & Discussions

### 3.1. Crystal structure

Figure 1(a) shows the powder x-ray diffraction patterns of YBa$_2$Cu$_{3-x}$Co$_x$O$_{7-\delta}$ with x = 0, 0.01, 0.03 and 0.05. The powder diffraction result of these samples showing major phase of Y:123. All the peaks were indexed on the basis of Y:123 structure. The crystal structure was found to be orthorhombic with Co substitution up to x=0.05. We have calculated the lattice parameters using

XRD patterns shown in Figure 1(a). The lattice parameters of $YBa_2Cu_{3-x}Co_xO_{7-\delta}$ (x = 0, 0.01, 0.03, 0.05) are shown in Table 2.

The lattice parameters remained almost unchanged but there is a minute elongation in lattice parameter 'a' with the addition of cobalt (Figure 1(b)). This may reflect that the radius of cobalt ion $Co^{3+}$ (0.69 Å) is close to the radius of copper ion $Cu^{2+}$ (0.87Å). The oxygen content was estimated as about 6.83 from the relation between the c- axis lattice parameter and oxygen content in Ref. [10]. The effect of this oxygen doping on the copper valence is shown in first column of Table 3. In the copper oxide superconductor parent materials, the $Cu^{2+}$ spin ½ ions (one unpaired electron per copper in the $d_{x2-y2}$ orbital) in this $CuO_2$ plane are ordered antiferromagnetically at a high temperature and the material is insulating. Superconductivity is induced when the number of electron in the $CuO_2$ plane is changed from one electron per copper site: i.e., the compounds are doped to make the formal copper valence different from $Cu^{2+}$ typically higher [11].

The minute elongation in lattice parameter 'a' is due to an increase in the oxygen content in the CuO chains. The cobalt doping may increase the incorporation of oxygen in the structure and thus the occupancy of the O(5) site and consequently an elongation in lattice parameter 'a'.

## 3.2. Scanning electron microscopy (SEM)

The scanning electron micrographs of cobalt substituted Y:123 samples are shown in Figure 2 (a to c). These electron micrographs reveal that the grain size in the cobalt doped sample is larger than that of undoped one.

In undoped sample the superconducting grains are well connected (see Figure 2(a)). When 0.01 mole of cobalt is added to Y:123 sample, few pores are found between regions of well connected grains. The pores/voids between the grains also increase with cobalt concentration up to 0.05 as shown in Figure 2(b) and Figure 2(c).

The above investigations revealed the fact that the superconducting grains are closely packed and pores between the grains are few (Figure 2(b) & 2(c)), the significant drops of Tc (onset) may be due to this non-homogeneous distribution of cobalt doping in the structure.

In order to further study of surface morphology we have obtained electron micrographs on these samples in backscattered mode (Figure 3(a) to Figure 3(c)). These electron micrographs in backscattered mode revealed the phase purity of the samples in local regions. In Figure 3(a) Y:123 phase is observed in all regions and white dots show the impurity phases. These impurity phases also increases with increasing concentration of Cobalt up to 0.05. The pores/voids are also observed to increase with increasing concentrations of cobalt (Figure 3(a) to Figure 3(c)).

### 3.3. Electrical resistance

The electrical resistance versus temperature variation, R (T) for the $YBa_2Cu_{3-x}Co_xO_{7-\delta}$ polycrystalline samples, with x=0, 0.01 and 0.05 is shown in Figure 4(a). All curves exhibit a transition to the superconducting state below the superconducting onset temperature Tc onset, from which the transition to zero resistance occurs. Such a critical temperature is related to the transition of isolated grains to the superconducting state in granular superconductors. The measurements were carried out on bar shaped samples with dimension $0.2 \times 0.14 \times 0.3 cm^3$. As evident from this figure, samples show suppression of Tc with increasing concentration of cobalt. The zero-resistance critical temperature decreases systematically with the increase in concentration of Cobalt (Figure 4(b)). The values of Tc (R=0) for the samples with Cobalt concentrations x = 0, 0.01, 0.03 and 0.05 are 90K, 75K, 70K and 60K respectively. This depression in Tc can be related to a decrease in hole concentration (hole filling) in the $CuO_2$ planes as a result of the cobalt substitution. When Co is substituted for Cu, $Cu^{2+}$ (0.87Å) ions are replaced by $Co^{3+}$ (0.69 Å) ions, each substitution of $Co^{3+}$ at $Cu^{2+}$ site fills holes and a decrease in hole concentration occurs on these planes . The effect of this cobalt substitution on copper valence is shown in Table 3.

Other probable reasons for this suppression of Tc are non-homogeneous distribution of cobalt doping in the structure [see section 3.2] and cooper pair breaking effect [12].

### 3.4. Transmission electron microscopy (TEM)

As a consequence of both the unavoidability and desirability of defects in the oxide superconductors, investigation by TEM in both imaging and selected area diffraction (SAD) mode is required to detect deviations from the average microstructure determined by SEM. Figure 5(a) shows a transmission electron micrograph exhibiting planar defects in local region of Co-doped Y:123 system and its corresponding SAD pattern along [001] direction is shown in Figure 5(b). These planar defects may serve as flux pinning centers.

The Co substituted compounds are found to contain secondary phases and the amount of these depending on the nominal x value (0 ≤ x ≤ 0.05). Figure 6(a) shows a micrograph of Co doped Y:123 system and corresponding SAD pattern along [010] direction is shown in Figure 6(b). These observations reveal the formation of Y:123 phase with small Co (0 ≤ x ≤ 0.01) substitution. For higher amount of Co substitution (0.03 ≤ x ≤ 0.05), impurity phases containing high Co are also identified.

Another interesting defect observed in Co-doped Y:123 material is the surface cleavage (planar defect). Internal structures with some periodicity terminate at the surface. Thus, the surface of as synthesised material may be considered in some senses to be a planar defect. At the surface, an atomic arrangement, which is different from the internal structure often, appears. Figure 7 shows such cleavages in Co-doped Y:123 system, shown by arrowhead in Figure.

## 3.5. Critical current density

The variation of the transport critical current density (Jc) of $YBa_2Cu_{3-x}Co_xO_{7-\delta}$ with different cobalt concentration revealed one order enhancement in Jc value as shown in Figure 8. In cobalt free sample critical current density (Jc) has been measured about $2.6 \times 10^2$ A/cm$^2$ but Jc=$3 \times 10^3$ A/cm$^2$ has been found for cobalt doped (x=0.05) sample. A serious handicap of the superconducting cuprates is the easy motion of their vortices, so that their critical current density tends to decrease dramatically when the temperature approaches Tc or when a high magnetic field is applied. People speculated that one could improve the critical current in crystals by introducing defects. A magnetic field penetrates a type-II superconductor in the form of vortices. Each vortex, carrying a flux quantum $\Phi$=h/2e, consists of a cylindrical core of radius $\xi$, the coherence length of the material, and a current circulating around the core out to a distance $\lambda$, the material penetration depth. When a current flows in the superconductor, the Lorentz force produces energy dissipation and consequently the disappearance of superconductivity. The mobility of the vortices can be minimized by introducing defects that pin the vortices. Thus the critical current density Jc depends on flux pinning of vortices. Defects such as stacking faults, pores/voids, screw dislocations, twin boundaries etc. are often found to be an effective flux pinning centers if their sizes are comparable to the coherence length ($\xi \sim$ 2-30 Å). In present Co doped Y-based system some useful defects such as planar defects, stacking faults and micro defects (twin, domains etc.) have already been observed in electron microscopic investigations. These defects may act as effective flux pinning centers and thus enhanced the $J_c$ values as observed in present case at the cost of $T_c$ suppression.

## 4. Conclusion

We have successfully prepared samples of type $YBa_2Cu_{3-x}Co_xO_{7-\delta}$ with x = 0, 0.01, 0.03 and 0.05 by standard ceramic method. The partial replacement of Cu by Co does not affect the orthorhombic structure of the Y:123 phase but there is a minute elongation in lattice parameter 'a'.

Surface morphology examination with SEM in scanning mode revealed the fact that the superconducting grains are closely packed and pores between the grains are in few numbers. These pores/voids between the grains increase with Cobalt concentration up to 0.05. On the other hand electron micrographs in backscattered mode revealed the phase purity of the samples in local regions. For lower amount of Co (x ≤ 0.01) small impurity phases have been found whereas for other higher amount of substitution of Co (x ≤ 0.05) more impurity phases were observed frequently.

The depression in transition temperature (Tc) as Co content increases has been explained due to the cooper pair breaking, hole filling in $CuO_2$ plane and non-homogeneous distribution of cobalt doping in the bulk materials. The TEM explorations also revealed the presence of defects such as stacking faults, planar defects and micro defects (twin, domains etc.) in the as synthesised doped samples. These defects may serve as flux-pinning centers. The critical current density $J_c$ increases as Co content increases. This may be due to the flux pinning effects such as stacking faults, planar defects and micro defects (twin, domains etc.).

## Acknowledgement

We are very much thankful to Prof. M. K. Mishra, vice-chancellor and Prof. U. D. Mishra, Head, Department of Physics, University of Lucknow for their encouragement and support for providing the facilities for material synthesis. We are also very much grateful to Prof. O.N. Srivastava and R. S. Tiwari, Department of Physics, Banaras Hindu University, Varanasi for fruitful and stimulated discussions.

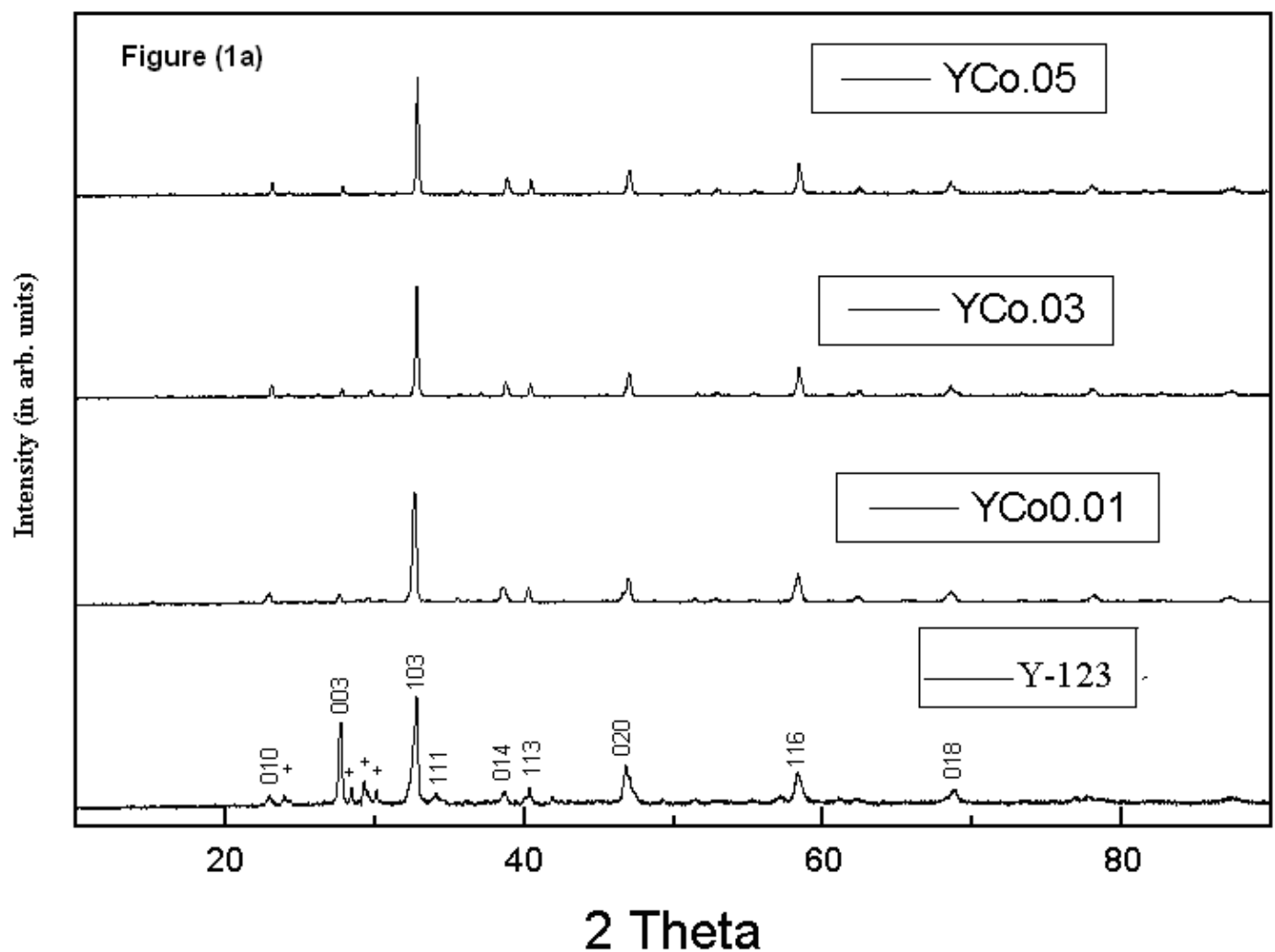

**Figure (1a):** X-ray powder diffraction patterns of YBa$_2$Cu$_{3-x}$Co$_x$O$_{7-\delta}$ system with x = 0.00, 0.01, 0.03 and 0.05.

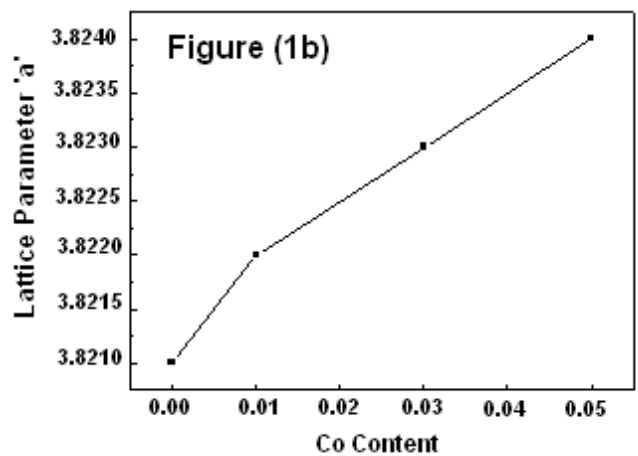

**Figure (1b):** The variation of lattice parameter 'a' versus Co-content shows minute elongation in lattice parameter 'a'.

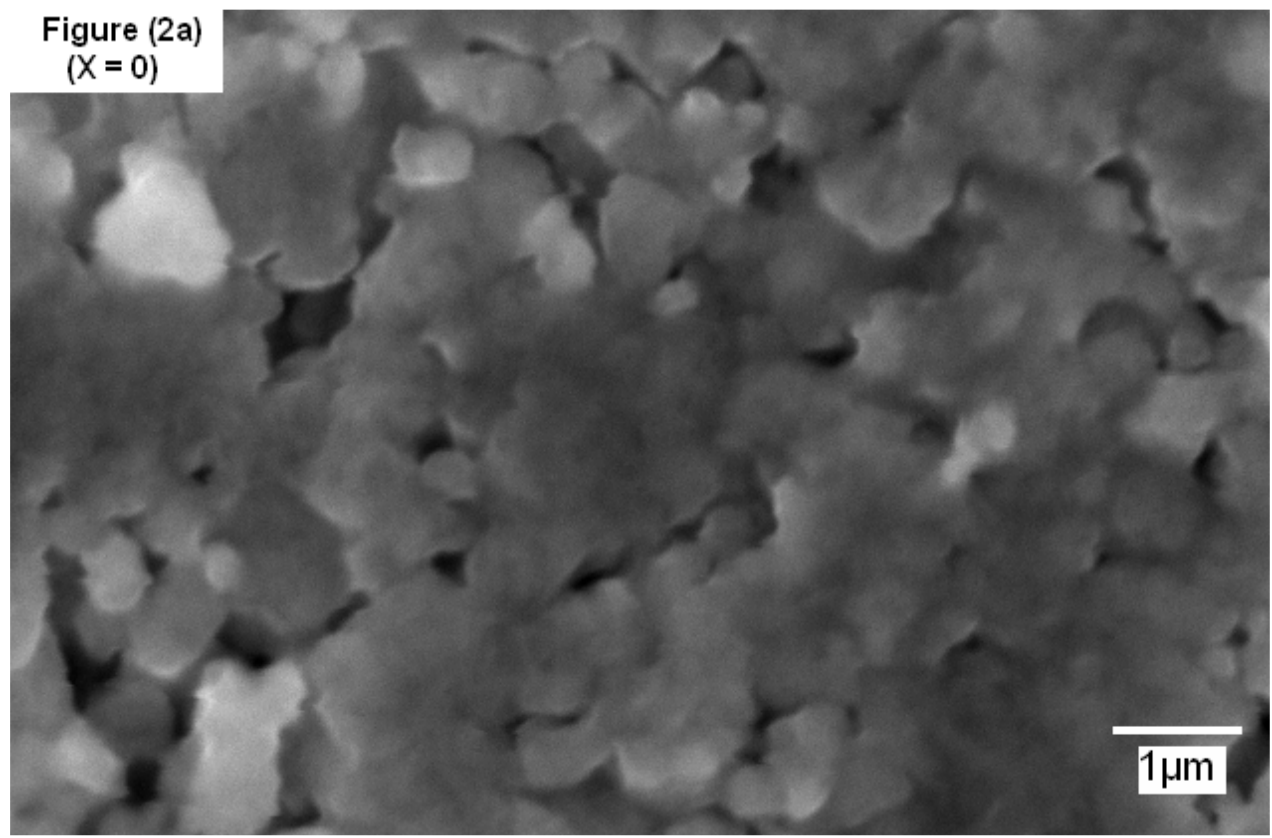

**Figure (2a):** Scanning electron micrograph of the surface of the bulk $YBa_2Cu_{3-x}Co_xO_{7-\delta}$ system with x = 0.00.

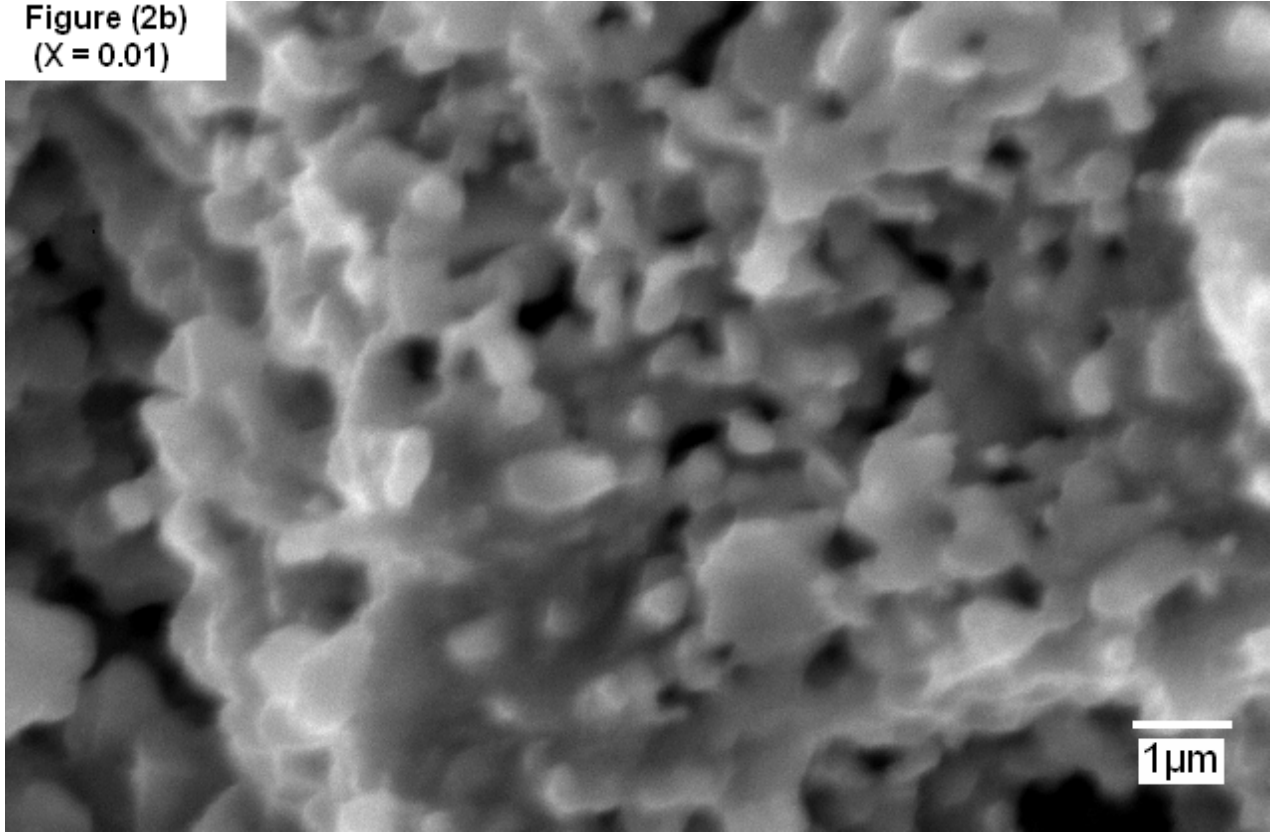

**Figure (2b):** Scanning electron micrograph of the surface of the bulk $YBa_2Cu_{3-x}Co_xO_{7-\delta}$ system with x = 0.01.

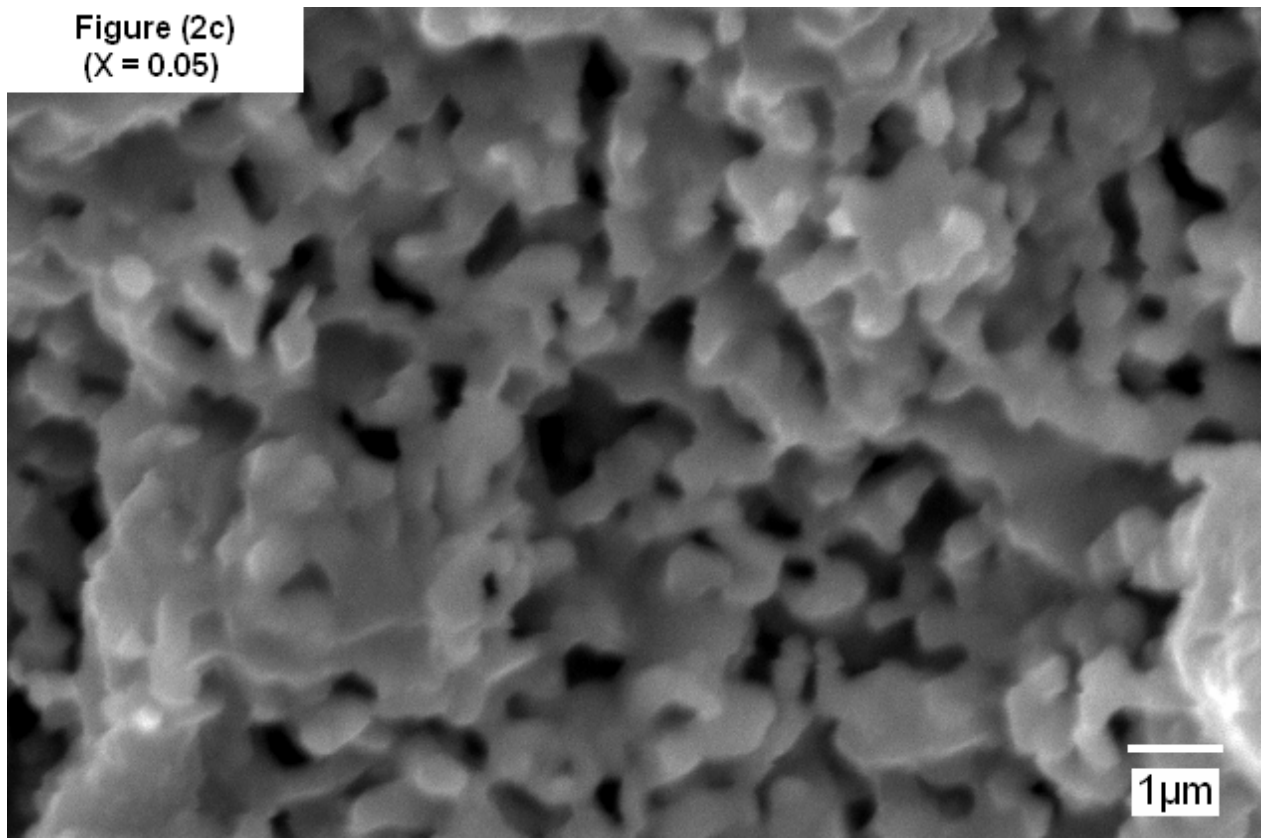

**Figure (2c):** Scanning electron micrograph of the surface of the bulk $YBa_2Cu_{3-x}Co_xO_{7-\delta}$ system with x = 0.05.

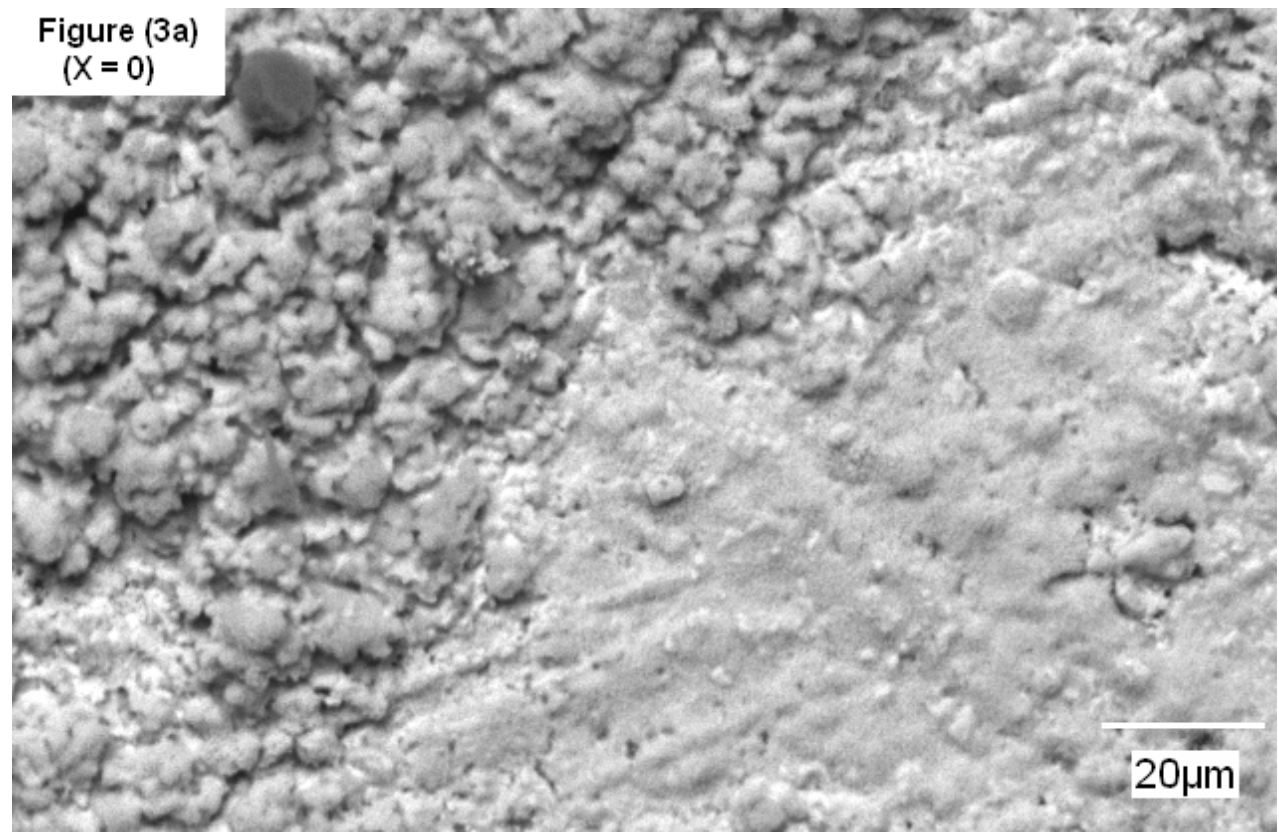

**Figure (3a):** SEM micrograph of $YBa_2Cu_{3-x}Co_xO_{7-\delta}$ with x = 0.00 taken in back scattered mode.

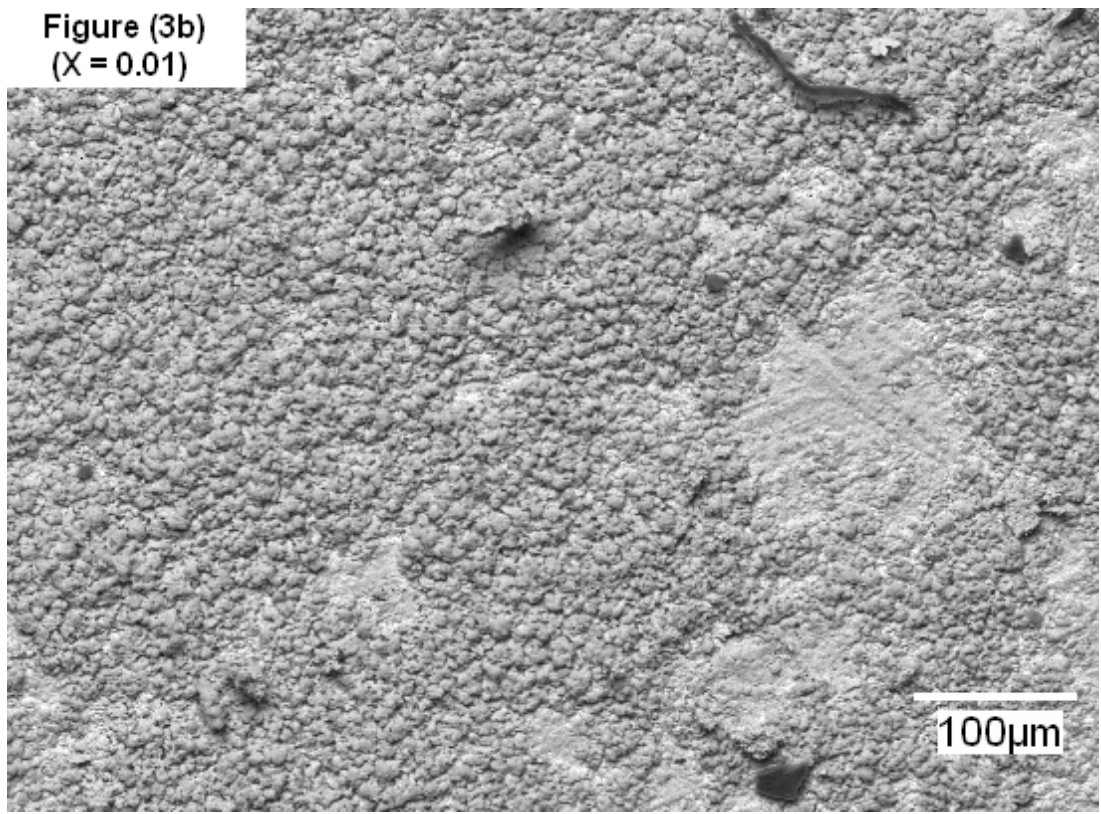

**Figure (3b):** SEM micrograph of YBa$_2$Cu$_{3-x}$Co$_x$O$_{7-\delta}$ with x = 0.01 taken in back scattered mode.

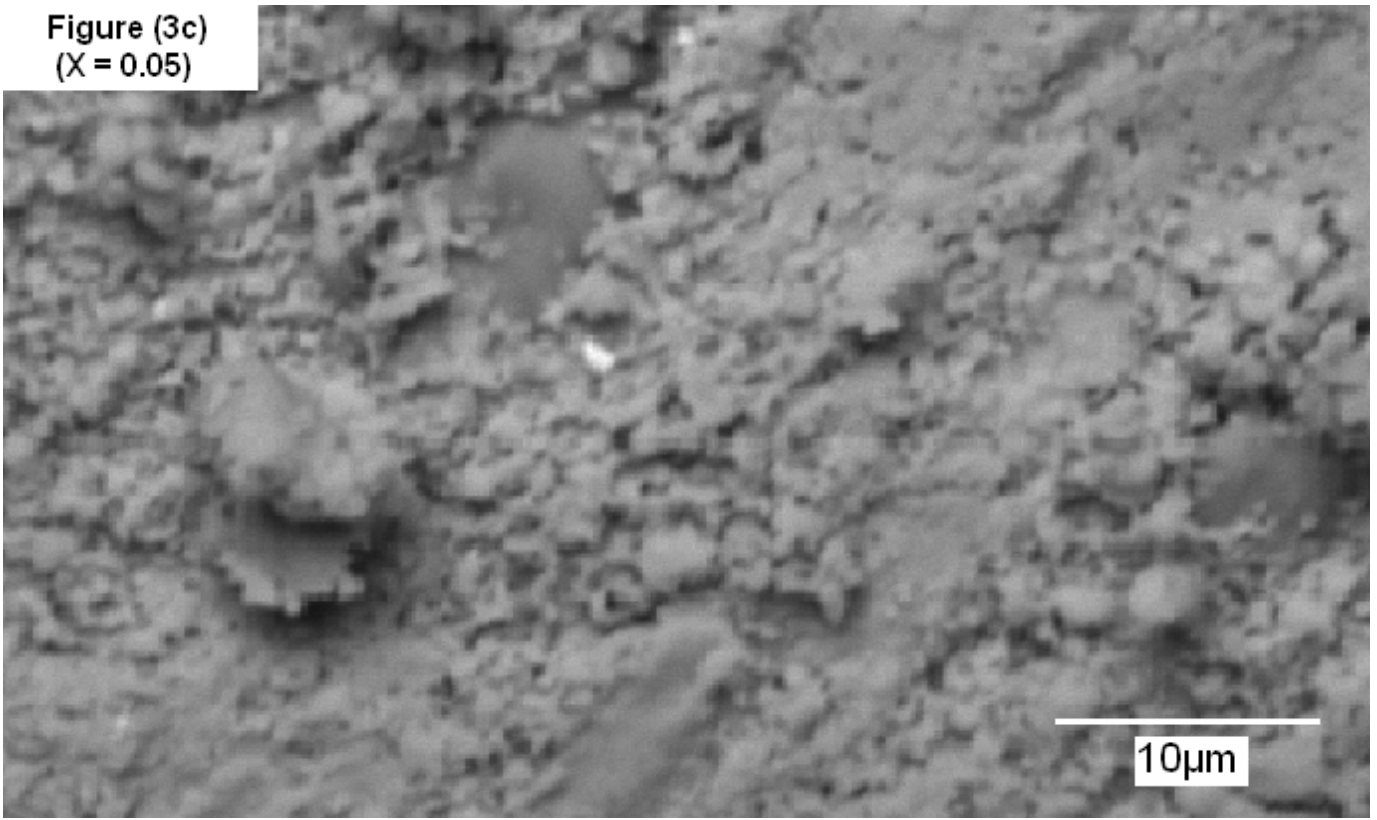

**Figure (3c):** SEM micrograph of $YBa_2Cu_{3-x}Co_xO_{7-\delta}$ with x = 0.05 taken in back scattered mode.

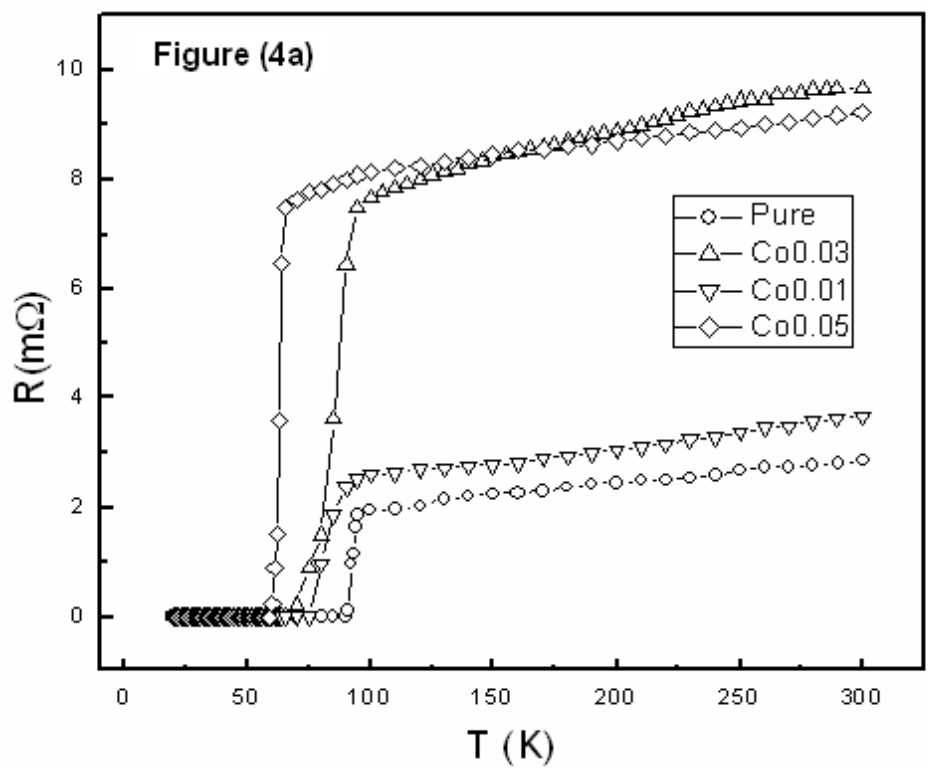

**Figure (4a):** Resistance (R) versus (T) plots of the samples $YBa_2Cu_{3-x}Co_xO_{7-\delta}$ for the various values of x.

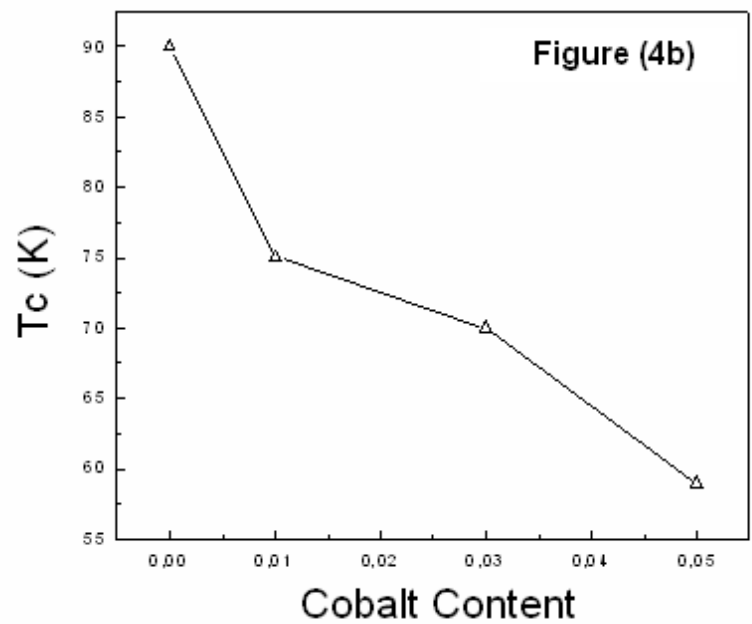

**Figure (4b):** The variation of transition temperature versus Co-content shows suppression of Tc with increasing Co concentration.

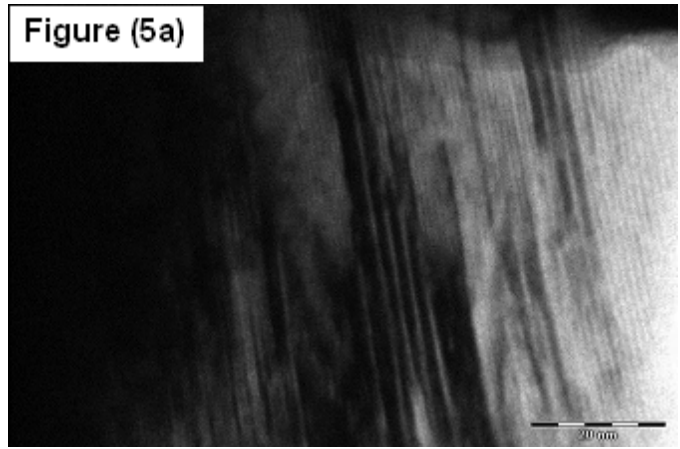

**Figure (5a):** Transmission electron micrograph showing planar defects in local region of Co doped Y:123.

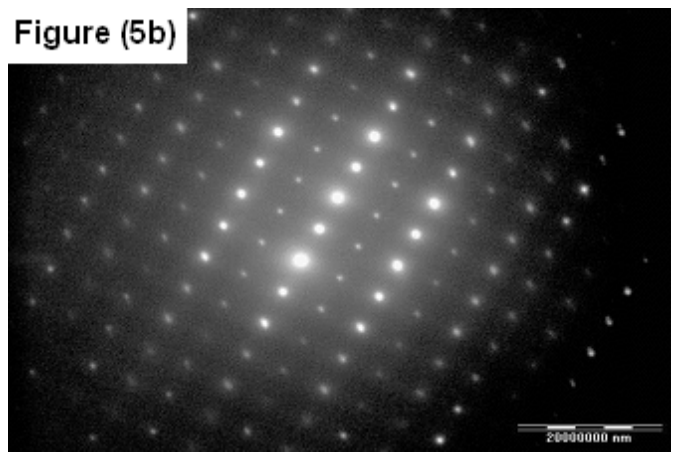

**Figure (5b):** Selected area electron diffraction (SAD) pattern corresponding to Figure (5a) along [001] direction of Co doped Y:123.

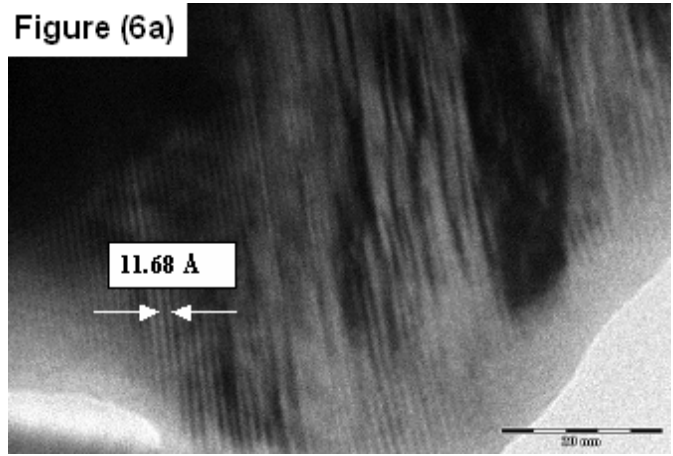

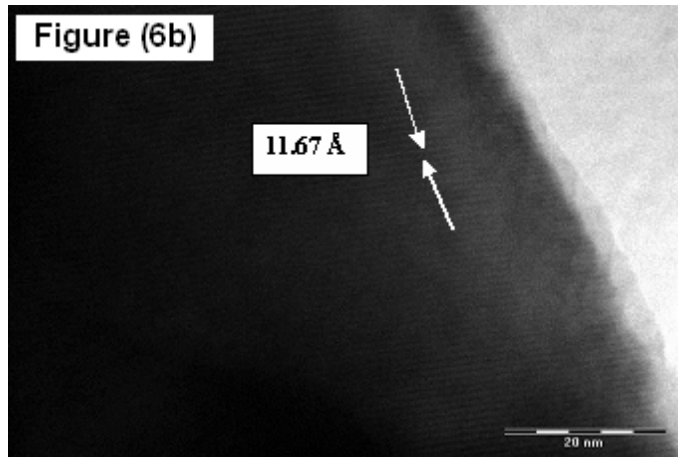

**Figure (6a) & (6b):** The direct lattice resolution of cobalt doped Y:123 system exhibiting the c periodicity

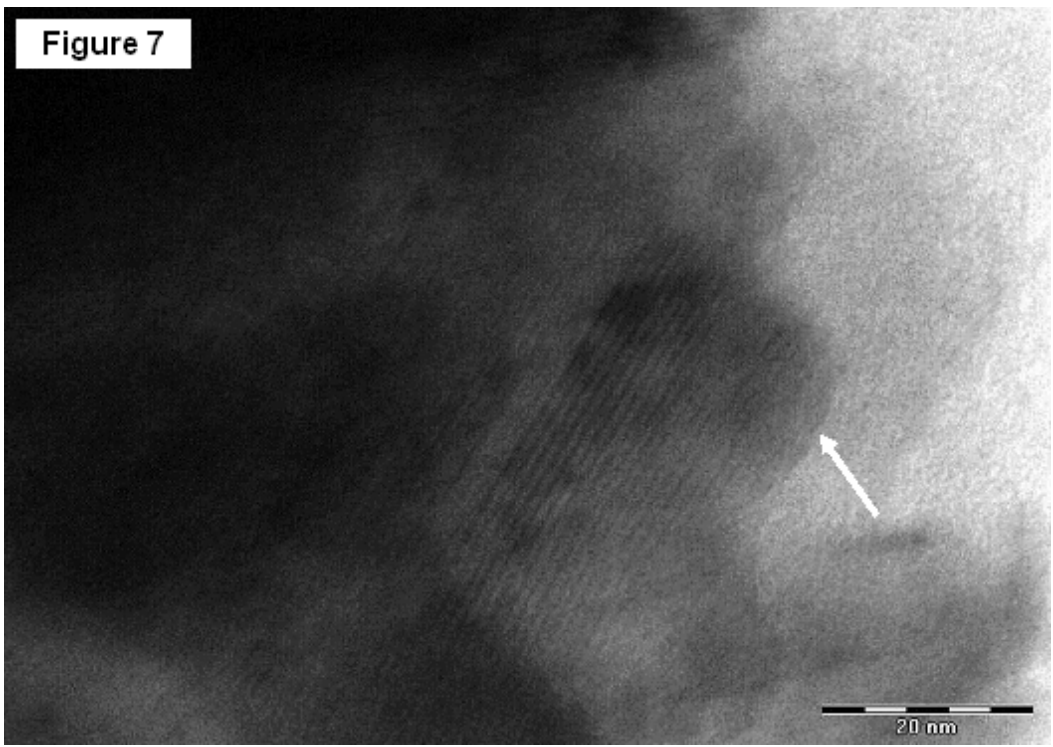

**Figure 7:** Structure image showing the cleavage surface structure of Co-doped Y:123 system.

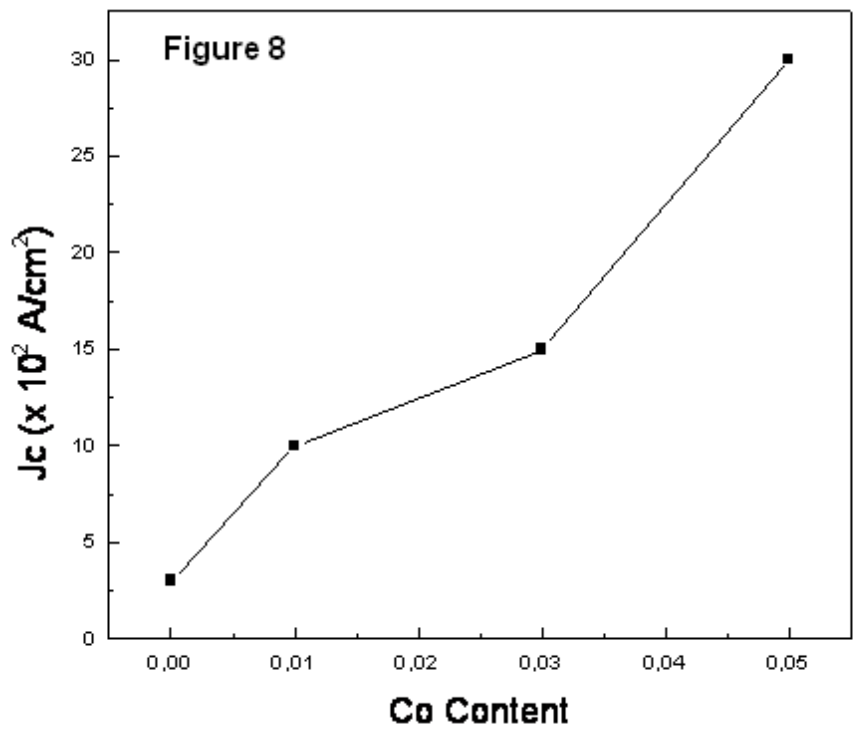

**Figure 8:** The variation of transport critical current density Jc versus Co-content revealing enhancement in critical current density with cobalt concentration.

**Table 1.** Nominal composition, sintering condition and annealing condition are shown for typical samples.

| Nominal composition of Co | Composition | Sintering conditions | Annealing conditions |
|---|---|---|---|
| X=0.00 | $YBa_2Cu_3O_{7-\delta}$ | $920^0C$ for 18 hr in air | $550^0C$ in $O_2$ for 13 hr at partial pressure $10^{-1}$ atm |
| X=0.01 | $YBa_2Cu_{2.99}Co_{0.01}O_{7-\delta}$ | $915^0C$ for 16 hr in air | $550^0C$ in $O_2$ for 14 hr at partial pressure $10^{-1}$ atm |
| X=0.03 | $YBa_2Cu_{2.97}Co_{0.03}O_{7-\delta}$ | $925^0C$ for 20 hr in air | $550^0C$ in $O_2$ for 12 hr at partial pressure $10^{-1}$ atm |
| X=0.05 | $YBa_2Cu_{2.95}Co_{0.05}O_{7-\delta}$ | $922^0C$ for 17hr in air | $550^0C$ in $O_2$ for 16 hr at partial pressure $10^{-1}$ atm |

**Table 2.** Lattice parameters of $YBa_2Cu_{3-x}Co_xO_{7-\delta}$ (x = 0 - 0.05).

| Nominal composition of Cobalt | Lattice parameters | | |
|---|---|---|---|
| | a (Å) | b (Å) | c (Å) |
| X=0.00 | 3.8210 | 3.870 | 11.6926 |
| X=0.01 | 3.8220 | 3.865 | 11.6908 |
| X=0.03 | 3.8230 | 3.862 | 11.6890 |
| X=0.05 | 3.8240 | 3.860 | 11.6883 |

**Table 3.** Effect of Oxygen doping and cobalt doping on copper valence.

| Effect of oxygen doping on copper valence | Effect of cobalt doping on copper valence after Oxygen doping | | |
|---|---|---|---|
| | Nominal composition of cobalt | | |
| | X=0.01 | X=0.03 | X=0.05 |
| $Y^{3+}Ba^{2+}_2Cu^{y+}_3O^{2-}_{6.83}$<br>$(3\times1)+(2\times2)+(3\times y)=(2\times6.83)$<br>**copper valence (y) = 2.22 (hole doping)** | $Y^{3+}Ba^{2+}_2Cu^{y+}_{2.99}Co^{3+}_{0.01}O^{2-}_{6.83}$<br>$(3\times1)+(2\times2)+(2.99\times y)+(3\times0.01)=(2\times6.83)$<br>**copper valence (y) = 2.2174 (hole filling)** | $Y^{3+}Ba^{2+}_2Cu^{y+}_{2.97}Co^{3+}_{0.03}O^{2-}_{6.83}$<br>$(3\times1)+(2\times2)+(2.97\times y)+(3\times0.03)=(2\times6.83)$<br>**copper valence (y) = 2.2121 (hole filling)** | $Y^{3+}Ba^{2+}_2Cu^{y+}_{2.95}Co^{3+}_{0.05}O^{2-}_{6.83}$<br>$(3\times1)+(2\times2)+(2.95\times y)+(3\times0.05)=(2\times6.83)$<br>**copper valence (y) = 2.207 (hole filling)** |